\documentclass[aps,prd,showpacs,amssymb,amsmath,nofootinbib]{revtex4}

\usepackage{amsthm}
\usepackage{amsfonts}
\usepackage{amscd}

\def\Mpl{\ensuremath{M_\textrm{Pl}}}
\def\odifone#1#2{\ensuremath{\frac{d #1}{d #2}}}
\def\Ztwo{\ensuremath{\mathbb{Z}_2}}
\def\Zfour{\ensuremath{\mathbb{Z}_4}}
\def\Mbare{\ensuremath{M_\textrm{bare}}}
\def\3by3Mat#1#2#3#4#5#6#7#8#9{\ensuremath{\begin{pmatrix}#1&#2&#3\\
                                                          #4&#5&#6\\
                                                          #7&#8&#9\end{pmatrix}}}
\def\2by2Mat#1#2#3#4{\ensuremath{\begin{pmatrix}#1&#2\\
                                                #3&#4\end{pmatrix}}}
\def\by32Mat#1#2#3#4#5#6{\ensuremath{\begin{pmatrix}#1&#2\\
                                                    #3&#4\\
                                                    #5&#6\end{pmatrix}}}
\def\Mat21#1#2{\ensuremath{\begin{matrix}#1\\
                                         #2\end{matrix}}}

\def\Dvev{\ensuremath{\langle\Phi\rangle}}
\def\Svev{\ensuremath{\langle\chi\rangle}}
\def\msol{\ensuremath{m_\textrm{sol}}}
\def\matm{\ensuremath{m_\textrm{atm}}}
\def\order#1{\ensuremath{\mathcal{O}(#1)}}
\def\curlN{\ensuremath{\mathcal{N}}}
\def\m1eff{\ensuremath{\widetilde{m}_1}}

\begin{document}


\title{Leptogenesis implications in models with Abelian family symmetry and one extra real Higgs singlet}
\author{Sandy~S.~C.~Law}
\email{slaw@physics.unimelb.edu.au}
\author{Raymond~R.~Volkas}
\email{rrv@physics.unimelb.edu.au}
\affiliation{School of Physics, Research Centre for High Energy
Physics,\\The University of Melbourne, Victoria 3010, Australia}
\date{\today}

\begin{abstract}
We show that the neutrino models, as suggested by Low \cite{Low:2005yc}, which have an additional Abelian family symmetry and a real Higgs singlet to the default see-saw do not hinder the possibility of successful thermal leptogenesis. For these models (neglecting radiative effects), we have investigated  the situation of strong washout in both the one-flavor approximation and when flavor effects are included. The result is that while such models predict that $\theta_{13}=0$ and that one light neutrino to be massless, they do not modify or provide significant constraints on the typical leptogenesis scenario where the final asymmetry is dominated by the decays of the lightest right-handed neutrinos.
\end{abstract}


\maketitle


\section{Introduction}\label{sec_intr}

The evidence for non-zero neutrino masses \cite{neutrinos_exp} and cosmological matter-antimatter asymmetry \cite{wmap} provides a strong indication for physics beyond the Standard Model (SM). An attractive way to explain these phenomena involves the inclusion of heavy right-handed (RH) Majorana neutrinos to the leptonic sector. As a result, tiny neutrino massess may be generated via the \textit{see-saw} mechanism \cite{seesaw_list} while the cosmic baryon asymmetry can be explained by \textit{thermal leptogenesis} \cite{Fukugita:1986hr}.\\

One particularly beautiful aspect of leptogenesis is that it offers a link between the baryon asymmetry and SM neutrino properties. The mechanism works by assuming that heavy RH neutrinos interact with the ordinary light neutrinos in a $CP$ and lepton number $(L)$ violating way in the early universe, and hence, generating an $L$ asymmetry which, given the right conditions ($100\lesssim T\lesssim 10^{12}$ GeV) \cite{KlinkMantonKRS}, would eventually be partially converted to a baryon number $(B)$ asymmetry by non-perturbative sphaleron processes. In addition, through the see-saw formula, relation of the mass scales of the light and heavy neutrino spectra can be established. Therefore, the requirement of successful leptogenesis naturally leads to limits on neutrino masses and mixings.\\

Given this intricate connection, it is of great interest to explore possible neutrino models that can give rise to successful leptogenesis. By the same token, it is intriguing to ask what are the implications of a given model, which is specifically designed to address a different issue, in the leptogenesis context. The aim of this paper is to conduct such studies on a specific class of see-saw neutrino models (as proposed by Low \cite{Low:2005yc}) that has an Abelian family symmetry and an extra real singlet in the Higgs sector. Because of the interplay between the new ingredients, these models predict a fully hierarchical light neutrino spectrum (ie. $m_1 = 0$), as well as, $\theta_{13}=0$ in the mixing matrix. More importantly, they contain \textit{fewer free parameters than the default see-saw}, which is the original motivation behind the previous work of Low. The analysis here will focus on the suitability of these models for leptogenesis in several of the typical setups.\\

Over the years, there has been a dramatic increase in the sophistication of the quantitative analysis of leptogenesis. Many previously neglected effects such as thermal corrections \cite{Giudice:2003jh}, different washout \cite{Luty:1992un,Plumacher:1996kc,Buchmuller:2002jk,Buchmuller:2002rq, Hambye:2003rt} and spectator processes \cite{spectator1, spectator2} and, above all, flavor effects \cite{Barbieri:1999ma, Endoh:2003mz, Fujihara:2005pv, Vives:2005ra, Nardi:2006fx, Abada:2006fw, Abada:2006ea, Blanchet:2006be} have been considered in recent analyses. Other variations to the general scheme, including resonant leptogenesis \cite{PilafUnderW}, asymmetry production dominated by the decays of the second lightest RH neutrinos \cite{Blanchet:2006dq} and models with more than three heavy RH neutrinos \cite{4RH_models}, have also received attention. However, for definiteness, in this paper we concentrate on the case of two or three heavy RH neutrinos (depending on which sub-class of Low's models we are considering) and that their mass spectrum is hierarchical, ie. $M_{1}\ll M_{2}\ll M_{3}$. Furthermore, we assume that the final $L$, and hence, $B$ asymmetry is predominantly due to the decays of $N_1$'s. The present investigation makes use of various existing results and attempts to draw comparsions where appropriate with the hope to deduce whether these specific neutrino models may favor certain regimes or predict significant deviation from the standard scenarios (for both the one-flavor and the multi-flavor cases). In most situations, we work in the ``strong washout'' regime although some discussions on the ``weak washout'' regime are also included.\\

With these motivations in mind, the paper is structured as follows. Section~\ref{sec_model} provides an outline of the neutrino models under consideration. It serves to highlight all the key features, as well as to recast them in a form suitable for later discussions. We also comment on the fine-tuning of parameters in these models. Section~\ref{sec_lep1} begins with a brief overview of the leptogenesis scenario in the one-flavor case. The aim is to highlight the important issues and parameters so that a comparsion between the default see-saw and our specific models can be made. In Section~\ref{sec_lep2}, the analysis is repeated for the case when flavor effects are included. A summary of results is presented in Section~~\ref{sec_sum}.


\section{See-saw models with Abelian family symmetry and an extra real Higgs singlet}\label{sec_model}

While the SM has been very successful in explaining the dynamics of sub-atomic particles, it is not without its shortcomings. A typical example is its inability to predict quark and lepton masses. Such parameters are put in by hand using data from experiments. Altogether, there are almost 20 free parameters in the minimal SM; and should neutrinos gain mass via the see-saw mechanism, even more parameters will be required. Therefore, from the model building point of view, it is natural to look for ways to reduce the number of parameters.\\

A first step towards this may be achieved by enlarging the symmetry in the leptonic sector so that the symmetry group becomes
\begin{equation}\label{equ_group}
 G = SU(3)_c \otimes SU(2)_L \otimes U(1)_Y \times G_{\textrm{family}},
\end{equation}
where $G_{\textrm{family}}$ is a leptonic family symmetry. But it has been shown in \cite{Low:2003dz} that models with an unbroken family symmetry and just the SM Higgs do not have more predictive powers than the default see-saw. Hence, one would need to expand the Higgs sector at the same time.\footnote{Alternatively, other symmetry breaking mechanisms could be used instead.} The exact nature of the coupling between the lepton and Higgs fields dictates the neutrino mixing matrix, $U_{\textrm{PMNS}}$, produced via family symmetries. In the work of Low \cite{Low:2005yc, Low:2004wx}, effects of Abelian family symmetries with additional Higgs singlets, doublets or triplets on certain features of the mixing matrix were studied. It was found that the simplest models that can predict $\theta_{13}=0$ contain one extra real Higgs singlet which transforms non-trivially in family space.\footnote{The CHOOZ experiment \cite{Apollonio:1999ae} indicates that $\theta_{13}$ is very small and, at best fit, it is taken to be zero.} The action of the family symmetry with one singlet can be summarised as follows. Suppose the leptonic Yukawa and mass terms are:\footnote{Note that it may be tempting to exclude the $\Mbare$ term in (\ref{equ_lag1}) and have RH neutrino massess generated by the $\chi$ term only. But it turns out that, without $\Mbare$, the solar mixing angle, $\theta_{12}$, is forced to be maximal by the symmetry (which is incompatible with the best fit data \cite{sol_data}).}
\begin{equation}\label{equ_lag1}
 \mathcal{L}_{\textrm{mass}} = -\overline{L}\,Y_l\,\Phi\, l_R
                               -\overline{L}\,Y_\nu\,\widetilde{\Phi}\,\nu_R
                               -\frac{1}{2}\overline{(\nu_R)^c}\,Y_{\chi}\,\chi\;\nu_R
                               -\frac{1}{2}\overline{(\nu_R)^c}\,M_\textrm{bare}\,\nu_R
                               + \textrm{h.c.},
\end{equation}
where $Y_l, Y_\nu, Y_\chi$ and $\Mbare$ are coupling matrices in the family basis. $L=(\nu_L, l_L)^T$ is the left-handed (LH) lepton doublet while $l_R$ and $\nu_R$ are the RH charged and neutral lepton singlets respectively. The Higgs fields are $\Phi$ (SM complex doublet) and $\chi$ (real singlet) with $\widetilde{\Phi}\equiv i\sigma_2 \Phi^*$. Family symmetry, $G_{\textrm{family}}$, demands that the full Lagrangian is invariant under the unitary transformations,
\begin{equation}\label{equ_transf}
 L\rightarrow S_L L,\quad l_R\rightarrow S_{l_R}l_R,\quad 
 \nu_R\rightarrow S_{\nu_R}\nu_R,\quad \Phi\rightarrow S_\Phi\Phi\quad\textrm{and}\quad
 \chi\rightarrow S_\chi\chi,
\end{equation}
in family space. These transformations restrict the coupling matrices to certain forms and thus it is possible to generate a mixing matrix with $U_{e3}=0$. Here, $U_{\textrm{PMNS}}$ is parametrised by \cite{Eidelman:2004wy}
\begin{equation}\label{equ_PMNS}
 U \equiv U_{\textrm{PMNS}}= 
 \3by3Mat{c_{12}c_{13}}{s_{12}c_{13}}{s_{13}\,e^{-i\delta}}
 {-s_{12}c_{23}-c_{12}s_{23}s_{13}\,e^{i\delta}}
 {c_{12}c_{23}-s_{12}s_{23}s_{13}\,e^{i\delta}}
 {s_{23}c_{13}}
 {s_{12}s_{23}-c_{12}c_{23}s_{13}\,e^{i\delta}}
 {-c_{12}s_{23}-s_{12}c_{23}s_{13}\,e^{i\delta}}
 {c_{23}c_{13}}\times D_\varphi,
\end{equation}
where $D_\varphi = \textrm{diag}(e^{i\varphi_1/2},e^{i\varphi_2/2},1)$ is the matrix containing all the Majorana phases while $\delta$ is the $CP$ violating Dirac phase, and $s_{mn}=\sin\theta_{mn}, c_{mn}=\cos\theta_{mn}$. Therefore, $U_{e3}=0$ corresponds to $\theta_{13}=0$ in this parametrisation. A possible representation of the sets of transformations that leads to this is shown in Table~\ref{table_1}. Their associated Abelian symmetry and coupling matrices are presented in Table~\ref{table_2}. Although the analysis is done assuming that there are three RH neutrinos, the desired property of $U_{e3}=0$ can also be generated by models with only two RH neutrinos. This is evident from the vanishing third column of $Y_\nu$ in Case 1 of Table~\ref{table_2}. Hence, in this case, the third RH neutrino is actually decoupled from the LH sector. If we remove it from the theory, Case 1 is reduced to Case 3 (see Table~\ref{table_1} and \ref{table_2}).\footnote{In fact the physics of Case 1 and 3 are equivalent, and henceforth, we shall not discuss Case 3.} It should be noted that these models keep the atmospheric mixing angle, $\theta_{23}$, and the solar mixing angle, $\theta_{12}$, as free parameters. Despite this and the more complicated Higgs sector, it has been shown that, overall, these models have less arbitrary parameters than the standard see-saw.\\

\begin{table}[t]
\begin{center}
\begin{tabular}{|c|c|c|c|c|}
\hline
 &$S_L=S_{l_R}$ &$S_{\nu_R}$ &$S_\Phi, S_\chi$ &$G_\textrm{family}$\\
\hline
1&$\3by3Mat{1}{0}{0}{0}{-1}{0}{0}{0}{-1}$
 &$\3by3Mat{1}{0}{0}{0}{-1}{0}{0}{0}{i}$
 &$\Mat21{\Phi\rightarrow\Phi,}{\chi\rightarrow -\chi}$
 &$\Zfour$\\
\hline
2&$\3by3Mat{1}{0}{0}{0}{-1}{0}{0}{0}{-1}$
 &$\3by3Mat{1}{0}{0}{0}{-1}{0}{0}{0}{1}$
 &$\Mat21{\Phi\rightarrow\Phi,}{\chi\rightarrow -\chi}$
 &$\Ztwo$\\ 
\hline
3&$\3by3Mat{1}{0}{0}{0}{-1}{0}{0}{0}{-1}$
 &$\2by2Mat{1}{0}{0}{-1}$
 &$\Mat21{\Phi\rightarrow\Phi,}{\chi\rightarrow -\chi}$
 &$\Ztwo$\\
\hline
\end{tabular}\caption{Diagonal representation of the transformations in family space that gives $U_{e3}=0$ for three or two (in Case 3) RH neutrinos.}
\label{table_1}
\end{center}
\end{table}
\begin{table}[t]
\begin{center}
\begin{tabular}{|c|c|c|c|c|}
\hline
 &$Y_l$ &$Y_\nu$ &$Y_\chi$ &$M_\textrm{bare}$\\
\hline
1&$\3by3Mat{\times}{0}{0}{0}{\times}{\times}{0}{\times}{\times}$
 &$\3by3Mat{\times}{0}{0}{0}{\times}{0}{0}{\times}{0}$
 &$\3by3Mat{0}{\times}{0}{\times}{0}{0}{0}{0}{\times}$
 &$\3by3Mat{\times}{0}{0}{0}{\times}{0}{0}{0}{0}$\\
\hline
2&$\3by3Mat{\times}{0}{0}{0}{\times}{\times}{0}{\times}{\times}$
 &$\3by3Mat{\times}{0}{\times}{0}{\times}{0}{0}{\times}{0}$
 &$\3by3Mat{0}{\times}{0}{\times}{0}{\times}{0}{\times}{0}$
 &$\3by3Mat{\times}{0}{\times}{0}{\times}{0}{\times}{0}{\times}$\\ 
\hline
3&$\3by3Mat{\times}{0}{0}{0}{\times}{\times}{0}{\times}{\times}$
 &$\by32Mat{\times}{0}{0}{\times}{0}{\times}$
 &$\2by2Mat{0}{\times}{\times}{0}$
 &$\2by2Mat{\times}{0}{0}{\times}$\\
\hline
\end{tabular}\caption{Coupling matrices generated by the transformations in Table~\ref{table_1} where ``$\times$'' denotes an arbitrary complex entry. Note that $Y_\chi$ and $M_\textrm{bare}$ are symmetric matrices.}
\label{table_2}
\end{center}
\end{table}
Furthermore, the texture zeros in the coupling matrices together with the see-saw formula
\begin{equation}\label{equ_seesaw}
 m_\nu \simeq -\widehat{m}_D M_R^{-1} \widehat{m}_D^{T},
\end{equation}
give rise to an $m_\nu$ (in any basis choice for $m_l$ and $M_R$) which has the following form:
\begin{equation}\label{equ_M_nu}
 m_\nu = \3by3Mat{a_1}{a_2 b_1}{a_2 b_2}
                 {a_2 b_1}{a_3 b_1^2}{a_3 b_1 b_2}
                 {a_2 b_2}{a_3 b_1 b_2}{a_3 b_2^2}, 
 \quad\textrm{where}\; a_1, a_2, a_3, b_1, b_2\in \mathbb{C}.
\end{equation}
In the above, $\widehat{m}_D=Y_\nu\Dvev, m_l=Y_l\Dvev$ and $M_R = Y_\chi\Svev + \Mbare$ where $\Dvev$ and $\Svev$ are the VEV of fields $\Phi$ and $\chi$ respectively. This structure of $m_\nu$ in (\ref{equ_M_nu}) predicts one of the Majorana mass eigenstates $(\nu)$ of the light neutrinos to be massless. For better illustration and subsequent discussions, it is convenient to rewrite the Lagrangian of (\ref{equ_lag1}) in the mass eigenbasis of the charged leptons and heavy RH Majorana neutrinos:
\begin{equation}\label{equ_lag2}
 \mathcal{L}_{\textrm{mass}} = -\frac{1}{2}\overline{N}\,D_M\,N
                               -\overline{l}\,h_l\,\Phi\, e
                               -\overline{l}\,h_\nu\,\widetilde{\Phi}\,N 
                               +\textrm{h.c.},
\end{equation}
where $l =(\nu_L',l_L')^T$ and $e$ are the charged lepton doublet and singlet respectively. We have defined the heavy Majorana neutrino field:\footnote{In general, $\nu_L'\neq\nu$ (the mass eigenstate for light neutrinos). $\nu_L'$ and $\nu_R'$ are new fields from the change of basis.} $N=(\nu_R'+\nu_R'^c)/\sqrt{2}$. The charged lepton mass matrix is given by $D_l\equiv\textrm{diag}(m_e,m_\mu,m_\tau)=h_l\Dvev$ while for the heavy neutrinos, the mass matrix is $D_M=\textrm{diag}(M_1,M_2,M_3)$. Although $m_\nu$ produced via the see-saw formula (\ref{equ_seesaw}) is unaffected by the basis change, in general, $h_\nu$ in (\ref{equ_lag2}) would have texture zeros different from $Y_\nu$ (in fact, the texture zeros would disappear in most cases; see Table~\ref{table_3}). Using $U$ from (\ref{equ_PMNS}) to diagonalise $m_\nu$ and choosing the sign convention that $D_m=-U^\dagger m_\nu U^*$, one gets $D_m=\textrm{diag}(m_1, m_2, m_3)$ with $m_1=0$ and $m_1<m_2<m_3$. Therefore, the light neutrino mass spectrum is fully hierarchical in these models. Relations between the $m_j$'s can be established by invoking the neutrino mixing data:\footnote{The inverted hierarchy case is in brackets.}
\begin{align}
 m_3^2-m_2^2 &= \Delta\matm^2\; (\Delta m_\textrm{sol}^2),\\
 m_2^2-m_1^2 &= \Delta\msol^2\; (\Delta m_\textrm{atm}^2-\Delta m_\textrm{sol}^2).
\end{align}
When $m_1=0$, one then gets
\begin{align}
 m_3^2 &= \matm^2\; (\Delta\matm^2),\\
 m_2^2 &= \Delta\msol^2\; (\Delta m_\textrm{atm}^2-\Delta m_\textrm{sol}^2),
\end{align}
where $\matm\equiv\sqrt{\Delta m_\textrm{atm}^2+\Delta m_\textrm{sol}^2}\approx 0.05$ eV.\footnote{$\Delta\matm^2=(2.6\pm 0.4)\times 10^{-3} \textrm{eV}^2$ \cite{atm_data, Fogli:2004as} and $\Delta\msol^2=(8.0^{+0.4}_{-0.3})\times 10^{-5} \textrm{eV}^2$ \cite{Fogli:2004as, sol_data}.} It should be noted that the conclusion in this section remains valid under one-loop renormalisation group running \cite{Low:2005yc}.
\begin{table}[t]
\begin{center}
\begin{tabular}{|c|c|c|c|}
\hline
 &1 &2 &3 \\
\hline
$Y_\nu$
 &$\3by3Mat{\times}{0}{0}{0}{\times}{0}{0}{\times}{0}$
 &$\3by3Mat{\times}{0}{\times}{0}{\times}{0}{0}{\times}{0}$
 &$\by32Mat{\times}{0}{0}{\times}{0}{\times}$\\
\hline
$h_\nu$
 &$\3by3Mat{\times}{\times}{0}{\times}{\times}{0}{\times}{\times}{0}$
 &$\3by3Mat{\times}{\times}{\times}{\times}{\times}{\times}{\times}{\times}{\times}$
 &$\by32Mat{\times}{\times}{\times}{\times}{\times}{\times}$\\
\hline
\end{tabular}\caption{The structure of the coupling matrices: $Y_\nu$ in the family basis and $h_\nu$ in the mass eigenbasis of the charged leptons and RH Majorana neutrinos. In general, ``$\times$'' denotes an arbitrary complex number.}
\label{table_3}
\end{center}
\end{table}
\subsection{Parameters fine-tuning}\label{subsec_finetune}

In order to make these models workable, the Higgs sector was expanded to accommodate the real singlet $\chi$. Its addition has inevitably introduced a new energy scale, $\Svev$, to the theory. Since $M_R = Y_\chi\Svev + \Mbare$ depends on this, it is essential to understand the implications of the scale of $\Svev$ in relation to other parameters in the model. The most general and renormalisable Higgs potential incorporating $\chi$ is given by
\begin{equation}\label{equ_V}
 V(\Phi, \chi)=\frac{1}{2}\mu_\phi (\Phi^\dagger\Phi)
              +\frac{1}{4}\lambda_\phi (\Phi^\dagger\Phi)^2
              +\frac{1}{2}\mu_\chi \chi^2
              +\frac{1}{4}\lambda_\chi \chi^4
              +\mu_{\phi\chi}(\Phi^\dagger\Phi)\chi^2,
\end{equation}
where $\mu_\phi$ and $\mu_\chi$ are in general functions of temperature $(T)$.
Also, if the potential is to be bounded from below, then $\mu_{\phi\chi} > -\sqrt{\lambda_\phi\lambda_\chi}/2$. From the see-saw mechanism, we expect that $\Svev\gg\Dvev$, and hence, $T_{c,\chi}\gg T_{c,\phi}\simeq\order{10^2}$ GeV, where $T_{c}$ denotes the critical temperature for symmetry restoration. In fact, one would want $T_{c,\chi} > T_\textrm{reh}$, the reheating temperature, so that any topological defects (domain walls) created by the spontaneous breaking of the $\Ztwo$ discrete symmetry of $\chi$ are eliminated via inflation. The required hierarchy, $\Svev\gg\Dvev$, is ensured (to tree-level) if $\mu_{\phi\chi}\rightarrow 0$. This can be seen from the tree-level minimum condition for (\ref{equ_V}):
\begin{align}
 \mu_\phi &= -\lambda_\phi\Dvev^2 - 2\mu_{\phi\chi}\Svev^2,\label{equ_min1}\\
 \mu_\chi &= -\lambda_\chi\Svev^2 - 2\mu_{\phi\chi}\Dvev^2.
\end{align} 
If $\order{\lambda_\phi}\simeq \order{\lambda_\chi}=\order{1}$, then $\mu_{\phi\chi}\rightarrow 0$ guarantees that $\mu_\phi$ remains at $\order{\Dvev^2}$. In this limit, it also means that $\Phi$ and $\chi$ fields are decoupled from each other.\\

Since, in a typical thermal leptogenesis analysis it is usually assumed that the reheating temperature $(T_\textrm{reh})$ after inflation is larger than the mass of the decaying heavy neutrino,\footnote{This is done so to minimise theoretical uncertainties although in certain regimes, this restriction may be relaxed without any appreciable change to the predictions. Here, the decaying neutrino refers to the one that dominates the $L$ asymmetry generation (usually $N_1$). Note that in the hierarchical limit for RH neutrinos, $N_3$ decays cannot be the dominant contribution to the final asymmetry \cite{DiBari:2005st}.} altogether, one has:
\begin{equation}\label{equ_T_rel}
 T_{c,\chi} > T_\textrm{reh}> M_{1\,\textrm{or}\,2}  \gg T_{c,\phi}.
\end{equation}
This relation implies that $Y_\chi$ must be suitably fine-tuned so that $Y_\chi\Svev\simeq \order{M_j}$. In other words, elements of $Y_\chi$ cannot be arbitrarily small, and as a result, radiative corrections to the potential $V(\Phi,\chi)$ due to $\chi$'s coupling to other fields in the model may destroy the hierarchy between $\Dvev$ and $\Svev$. Nonetheless, for the purpose of this paper, we shall work in the assumption that such stability problem and any additional fine-tuning of the framework will be inconsequential to our main discussion.


\section{Implications in leptogenesis within the one-flavor approximation}\label{sec_lep1}

In order to understand the potential implications of our specific models on leptogenesis predictions, it is essential to get oriented in the standard scenario. The key relation that captures the dependence of the predicted baryon to photon number ratio $(\eta_B)$ at recombination time on the elements of thermal leptogenesis can be written as \cite{Buchmuller:2004tu}
\begin{align}
 \eta_B &= \tilde{d} \sum_{j=1}^{3} \varepsilon_j\, \kappa_j^\textrm{f}\,, \nonumber\\
        &\simeq \tilde{d}\, \varepsilon_1\, \kappa_1^\textrm{f}\,,\qquad (N_1\textrm{-dominated case})
        \label{equ_etaB}
\end{align}
where $\tilde{d}$ is the dilution factor that accounts for the partial conversion of the generated $\eta_{B-L}$ into $\eta_B$ through sphaleron processes, as well as the increase of photon number per comoving volume from the onset of leptogenesis to recombination; $\varepsilon_j$ measures the $CP$ asymmetry in the decays of $N_j$:
\begin{equation}\label{equ_cp1}
 \varepsilon_j = 
   \frac{\Gamma(N_j\rightarrow l\Phi)-\Gamma(N_j\rightarrow \overline{l}\Phi^\dagger)}
        {\Gamma(N_j\rightarrow l\Phi)+\Gamma(N_j\rightarrow \overline{l}\Phi^\dagger)}
        \equiv
        \frac{\Gamma_j - \overline{\Gamma}_j}{\Gamma_j + \overline{\Gamma}_j}\,,
\end{equation}
while $\kappa_j^\textrm{f}$ represents the (final) efficiency factor for $B-L$ production from $N_j$ decays which takes into account the initial conditions and the dynamics of particle interactions in the leptogenesis era, in particular, the interplay between decays, inverse decays and $\Delta L\neq 0$ scatterings. Hence, there are potentially three places where our models may modify the overall $\eta_B$ prediction.

\subsection{Dilution factor}\label{subsec_dilution}

To track the time evolution of the number density of a quantity, $X$ (eg. $N_j$, $B-L$ or $B$), in an expanding universe, it is more convenient to consider the number of particles, $\curlN_X(t)$, in a portion of comoving volume $(R^3(t))$ that contains one photon at time $t' \ll t_\textrm{lepto}$ \footnote{$t_\textrm{lepto}$ denotes the time at the onset of leptogenesis.} than the conventional, $n_X(t)$. We choose the normalization for $R^3(t)$ such that in relativistic thermal equilibrium, it contains on average $\curlN_{N_j}^{\textrm{eq}}(t\ll t_\textrm{lepto})=1$ heavy RH neutrino. Hence, for the baryon asymmetry at recombination time, we have
\begin{equation}
 \eta_B = \frac{\curlN_B^\textrm{f}}{\curlN_\gamma(t_\textrm{rec})}\,,
\end{equation}
where $\curlN_B^\textrm{f}$ denotes the final $B$ value after leptogenesis and sphalerons. In the simplest case of constant entropy $(s)$ and assuming standard photons production from $t'$ to $t_\textrm{rec}$, one has 
\begin{equation}\label{equ_3_4}
 \curlN_\gamma(t_\textrm{rec})=
 \frac{\curlN_\gamma(t_\textrm{rec})}{\curlN_\gamma(t')}
  = \frac{4}{3}\times \frac{g^*_s(t')}{g^*_s(t_\textrm{rec})}
  = \frac{4}{3}\times \frac{434/4}{43/11}\approx 37. 
\end{equation}
The pre-factor of 4/3 originates from our choice of normalisation. $g^*_s(t)$ is the relativistic degrees of freedom at time $t$. In (\ref{equ_3_4}), we have already assumed the $N_1$-dominated scenario with $g^*_s(t')$ taken to be the degrees of freedom from all SM particles and $N_1$ only. Up to this point, the calculation is as per usual because the only new particle in our models is the physical $H_\chi$ which gains mass at a very high energy $(T_{c,\chi})$, and at time $t'$, it is non-relativistic.\footnote{In general, $M_{H_\chi}$ depends on $\lambda_\chi$ which is unknown. But as in Sec.~\ref{subsec_finetune}, we assume that $\order{\lambda_\chi}\simeq\order{\lambda_\phi}=\order{1}$, which is not unreasonable if one expects to find the Higgs $(H_\phi)$ at TeV scale. As a result, this implies that $M_{H_\chi}\simeq\order{\Svev}$ is very heavy. Moreover, even if $H_\chi$ is included in the calculation of $g^*_s(t')$, it will only change (\ref{equ_3_4}) by less than 2\%.}\\

Another element that contributes to the dilution factor comes from the imperfect conversion of $\curlN_{B-L}^\textrm{f}$ into $\curlN_{B}^\textrm{f}$. The simplified picture which assumes that sphaleron processes are only active after the leptogenesis era, gives a sphaleron conversion factor of $a_\textrm{sph}=\curlN_{B}^\textrm{f}/\curlN_{B-L}^\textrm{f} = 28/79$ \cite{YKS_HarvTurn} in the SM with one Higgs doublet.\footnote{This factor is somewhat different if electroweak sphalerons remain in equilibrium until slightly after $T_{c,\phi}$ \cite{Laine:1999wv}. Note that this case is highly probable because the electroweak phase transition seems to be not strongly first order.} In a more thorough analysis when the combined effect of all spectator processes \cite{spectator1, spectator2} in the plasma  (eg. Yukawa interactions, QCD and electroweak sphalerons) is taken into account, the resultant value receives a 20\% to 40\% enchancement or suppression \cite{spectator2} depending on the specific leptogenesis temperature $T_\textrm{lepto}$ assumed.\footnote{The temperature region which $T_\textrm{lepto}$ falls into also dictates the importance of flavors in leptogenesis. We defer our comments on this to Sec.~\ref{sec_lep2}.} This change comes about because chemical potentials of particles in thermal equilibirum are modified during leptogenesis, and not all of these potentials are independent due to SM, Yukawa and sphaleron interactions \cite{YKS_HarvTurn}. As a result, washout processes\footnote{These are the same washout processes that determine the efficiency factor $\kappa_j^\textrm{f}$. We include these effects here rather than in Sec.~\ref{subsec_eff} because they are related to spharlerons. But in cases when flavor effects are important, it would be more natural to incorporate them in the discussion of $\kappa_{j\alpha}^\textrm{f}$ (the flavor dependent efficiency factor).} that control $\curlN_{B-L}^\textrm{f}$ get enhanced or suppressed, leading to an overall change in $a_\textrm{sph}$. The dependence on temperature is originated from the fact that an increasing number of Yukawa or sphaleron processes comes into equilibrium as $T$ decreases.\\

Applying these ideas to our models, it is not hard to see that $a_\textrm{sph}$ receives no modification from the existence of $H_\chi$. This is because none of these interactions: $H_\chi\leftrightarrow N_j N_j$, $N_j N_j\leftrightarrow N_k N_k$ (via $H_\chi$) changes $\curlN_L$ or $\curlN_\phi$ in the plasma, and for most temperatures, they are not in equilibrium anyway due to the heaviness of $H_\chi$ and $N_j$. Furthermore, even if $H_\chi$ is light, the process
$H_\chi H_\chi\leftrightarrow \Phi\,\Phi^\dagger$ which can change $\curlN_\phi$ (and hence $\curlN_L$) is impotent since $\mu_{\phi\chi}\rightarrow 0$. Therefore, our specific models do not change the prediction of the dilution factor $\tilde{d}$ with respect to the standard see-saw setup no matter whether spectator processes are included.

\subsection{$CP$ asymmetry}\label{subsec_cp}

While dilution $\tilde{d}$ and efficiency $\kappa_j^\textrm{f}$ (see Sec.~\ref{subsec_eff}) govern the portion of the generated asymmetry that would survive after the entire process, it is the $CP$ and $L$ violating decays of the heavy neutrino ($N_j\rightarrow l\Phi$ or $\overline{l}\Phi^\dagger$) that give rise to such asymmetry in the first place. This quantity is defined as in (\ref{equ_cp1}) and its value is controlled by the so-called \textit{see-saw geometry} \cite{DiBari:2005st}. Note that at tree level
\begin{equation}\label{equ_tree_decay}
 \Gamma_j = \overline{\Gamma}_j = \frac{(h_\nu^\dagger h_\nu)_{jj}}{16\pi}M_j,
\end{equation}
where we have used the notation of (\ref{equ_lag2}). Thus, no asymmetry is generated at tree level. The leading contribution to $\varepsilon_j$ comes from the interference of the one-loop vertex and self-energy corrections with the tree level coupling. A perturbative calculation of this yields \cite{Covi:1996wh,CPasymmetry2}
\begin{equation}\label{equ_cp2}
 \varepsilon_j = \frac{1}{8\pi}\sum_{k\neq j} 
              \frac{\textrm{Im}\left[(h_\nu^\dagger h_\nu)^2_{jk}\right]}
               {(h_\nu^\dagger h_\nu)_{jj}}
             \left\{f_V\left(\frac{M_k^2}{M_j^2}\right)+f_S\left(\frac{M_k^2}{M_j^2}\right)
             \right\},
\end{equation} 
where $f_V(x)$ and $f_S(x)$ are given by
\begin{equation}
 f_V(x) = \sqrt{x}\left[1-(1+x)\ln\left(\frac{1+x}x{}\right)\right]\quad\textrm{and}\quad
 f_S(x) = \frac{\sqrt{x}}{1-x}\,,
\end{equation}
which denote the vertex and self-energy contributions respectively.

For $j=1$ and in the limit of hierarchical RH neutrino with $M_1\ll M_{2,3}$ (ie. $x\gg 1$), we have
\begin{equation}
 f_V(x)+f_S(x)\simeq -\frac{3}{2\sqrt{x}}\,.
\end{equation}
Therefore, the $CP$ asymmetry for $N_1$ decays in this limit is given by
\begin{align}
 \varepsilon_1 &\simeq -\frac{3 M_1}{16\pi (h_\nu^\dagger h_\nu)_{11}}\sum_{k\neq 1} 
             \textrm{Im}\left[(h_\nu^\dagger h_\nu)^2_{1k}\right]
             \frac{1}{M_k}\,,\label{equ_cp3a}\\
            &= -\frac{3 M_1}{16\pi (h_\nu^\dagger h_\nu)_{11}}\textrm{Im}
            \left[(h_\nu^\dagger h_\nu D_M^{-1} h_\nu^T h_\nu^*)_{11}\right].\label{equ_cp3}
\end{align} 
It is customary to rewrite (\ref{equ_cp3}) in terms of a complex orthogonal matrix $(\Omega)$ whose relation to the Yukawa coupling matrix $h_\nu$ is given by \cite{Casas:2001sr}
\begin{equation}\label{equ_omega1}
 \Omega = \Dvev\, D_m^{-1/2}\, U^\dagger\, h_\nu\, D_M^{-1/2} 
        \equiv D_m^{-1/2}\, U^\dagger\, m_D\, D_M^{-1/2}\quad \textrm{with}\quad \Omega^T\Omega=I,
\end{equation} 
where all symbols are defined as in Sec.~\ref{sec_model} with matrix $B=A^{-1/2}$ defined by $B^2=A^{-1}$ and $AA^{-1}=I$. In deriving (\ref{equ_omega1}), we have used the see-saw relation (\ref{equ_seesaw}) in the basis where the charged lepton and heavy neutrino mass matrices are real and diagonal (ie. $m_\nu \simeq -m_D D_M^{-1} m_D^T$) and the definition of $D_m =-U^\dagger m_\nu U^*$. Using (\ref{equ_omega1}) in (\ref{equ_cp3}) and after some manipulations, one gets \cite{Casas:2001sr}
\begin{align}
 \varepsilon_1 &\simeq \frac{3 M_1}{16\pi\Dvev^2}\frac{\sum_k m_k^2\, \textrm{Im}(\Omega_{k1}^2)}
             {\sum_k m_k\, |\Omega_{k1}^2|}\,,\label{equ_omega_1}\\
           &=\frac{3 M_1 \matm}{16\pi\Dvev^2}\,\beta(m_1,\m1eff,\Omega_{k1}^2)\,,\label{equ_cp4}
\end{align}
where we have introduced: the effective neutrino mass, $\m1eff$ and a dimensionless quantity, $\beta(m_1,\m1eff,\Omega_{k1}^2)$, which are defined as \cite{Plumacher:1996kc, DiBari:2005st}
\begin{align}
 \m1eff &\equiv \frac{(m_D^\dagger m_D)_{11}}{M_1}\,,\label{equ_m1eff}\\
\intertext{and}
  \beta(m_1,\m1eff,\Omega_{k1}^2) 
        &\equiv \frac{\sum_k m_k^2\, \textrm{Im}(\Omega_{k1}^2)}
             {\matm \sum_k m_k\, |\Omega_{k1}^2|}\,,\label{equ_beta1}\\
        &=\frac{\sum_k m_k^2\, \textrm{Im}(\Omega_{k1}^2)}
             {\matm\, \m1eff}\,,\label{equ_beta2}
\end{align}
respectively. Note that one can easily get to (\ref{equ_beta2}) by using relation (\ref{equ_omega1}) in (\ref{equ_m1eff}).\\

It will become apparent later (see Sec.~\ref{subsec_eff}) that the value of $\m1eff$ plays a crucial role in governing the washout regime that is relevant, as well as determining the final efficiency factor. Suppose that $\m1eff$ and $M_1$ have been fixed due to other considerations (eg. washout regime selected or potency of flavor effects), expression (\ref{equ_cp4}) then implies that the size of the $CP$ asymmetry is controlled only by $m_1$ and the configuration of the three $\Omega_{k1}^2$'s (which are all neutrino model dependent parameters). Although these $\Omega_{k1}^2$'s are generally complex, the orthogonality condition $(\sum_k \Omega_{k1}^2=1)$ means that only 3 real independent parameters are needed to specify them. Moreover, since the numerator of (\ref{equ_beta2}) depends on the imaginary part of $\Omega_{k1}^2$, only 2 of these 3 parameters will manifest itself in $\beta(m_1,\m1eff,\Omega_{k1}^2)$ when $\m1eff = \sum_k m_k |\Omega_{k1}^2|$ is a constant. This can be made clear if we define $\Omega_{k1}^2=X_{k}+iY_{k}$ and use the orthogonality condition to re-write (\ref{equ_beta2}) as
\begin{equation}\label{equ_beta3}
 \beta(m_1,\m1eff,\Omega_{k1}^2) = \frac{Y_2(m_2^2-m_1^2)+Y_3(m_3^2-m_1^2)}{\matm\,\m1eff}\,.
\end{equation}
An important question to ask though is which sets of $\Omega_{k1}^2$'s will yield a maximum $CP$ asymmetry given fixed values for $m_1$ and $\m1eff$. By the same token, it is useful to ascertain the upper bound\footnote{Without loss of generality, we may adopt the convention that the $CP$ asymmetry is positive and hence the lower bound is given by $0 \leq\beta(m_1,\m1eff,\Omega_{k1}^2)$.} for $\beta(m_1,\m1eff,\Omega_{k1}^2)$ for a specific class of neutrino models. For this purpose, it is convenient to break up the dependence on $m_1, \m1eff$ and $\Omega_{k1}^2$ and introduce an effective leptogenesis phase $\delta_L$ \cite{DiBari:2005st,Asaka_Hamaguchi, Buchmuller:2003gz, Davidson:2002qv}
\begin{align}
 \beta(m_1,\m1eff,\Omega_{k1}^2) 
 &=\beta_\textrm{max}(m_1,\m1eff)\sin \delta_L(m_1,\m1eff,\Omega_{k1}^2),\label{equ_beta4}\\
 &=\beta_1(m_1)\beta_2(m_1,\m1eff)\sin \delta_L(m_1,\m1eff,\Omega_{k1}^2),\label{equ_beta5}
\end{align}
where $\beta_\textrm{max}\leq 1$ represents the maximal value for $\beta$ given a particular $m_1$ and $\m1eff$. Clearly, the $CP$ asymmetry is largest when the configuration of $\Omega_{k1}^2$ leads to $\sin \delta_L=1$. The upper bound for $\beta$ when only $m_1$ is fixed is denoted by function $\beta_1$ while $\beta_2$ is a correction to $\beta_1$ when $\m1eff$ is also fixed at a given finite value. By analysing (\ref{equ_beta3}), it can be shown that \cite{DiBari:2005st, Davidson:2002qv}
\begin{equation}\label{equ_b_1}
 \beta_1(m_1) = \frac{m_3-m_1}{\matm}.
\end{equation}
To get $\beta_2$, one observes that for a generic $\Omega$ matrix, a configuration that maximises $\beta$ while keeping $\m1eff$ the same is achieved when $\Omega_{21}^2=0$ \cite{DiBari:2005st}. This allows one to rewrite (\ref{equ_beta3}) as
\begin{align}
 \beta &= \frac{Y_3(m_3^2-m_1^2)}{\matm\,\m1eff}\,,\label{equ_beta6}\\
       &= \frac{Y_{3\textrm{m}}\sin\delta_L (m_3^2-m_1^2)}{\matm\,\m1eff}\,,\label{equ_beta7}
\end{align}
where $Y_{3\textrm{m}}$ is the maximum value of $Y_3$ when $\Omega_{21}^2=0$. By putting (\ref{equ_beta5}), (\ref{equ_b_1}) and (\ref{equ_beta7}) together, one gets
\begin{equation}\label{equ_b_2}
 \beta_2 
 = \frac{\beta}{\beta_1\,\sin\delta_L} = \frac{m_1+m_3}{\m1eff}\,Y_{3\textrm{m}}.
\end{equation}
In general, $Y_{3\textrm{m}}$ will depend on the light neutrino masses. For a fully hierarchical neutrino spectrum ($m_1=0$) like our models with family symmetry, we have $\beta_1=m_3/\matm$ and $\beta_2=1$ (with $X_2=Y_2=X_3=0$ and $Y_{3\textrm{m}}=\m1eff/m_3$).\footnote{For normal heirarchy, $m_3=\matm$ by definition and so $\beta_1=1$.}\\

Using the above decomposition of the $CP$ asymmetry, it is then very easy to deduce the specific predictions of our models using the structure of $h_\nu$. Starting from the $h_\nu$'s given in Table~\ref{table_3},\footnote{We shall ignore Case 3 for it is effectively identical to Case 1.} we first note that for Case 2, the corresponding $\Omega$ matrix is arbitrary since all entries of $h_\nu$ in this case are unconstrained by the symmetry. As a result, from the above general analysis, it is clear that Case 2 certainly has enough parameter freedom to create a $CP$ asymmetry for the leptogenesis purpose, and in fact it predicts no additional restrictions other than those implied by the model independent analysis.\\

The situation for Case 1 is slightly different though due to the presence of texture zeros. For the purpose of working out $\Omega$, we replace all texture zeros in $h_\nu$ by an infinitesimal parameter $\varrho$. Hence, we have for this case
\begin{equation}\label{equ_varrho}
 h_\nu = \3by3Mat{\times}{\times}{\varrho}{\times}{\times}{\varrho}{\times}{\times}{\varrho}.
\end{equation}
Applying this in (\ref{equ_omega1}) yields an orthogonal matrix of the following form:
\begin{equation}\label{equ_omega2}
 \Omega=\3by3Mat{\propto \frac{1}{\sqrt{m_1 M_1}}}
 {\propto \frac{1}{\sqrt{m_1 M_2}}}{\propto \frac{\varrho}{\sqrt{m_1 M_3}}}
 {\propto \frac{1}{\sqrt{m_2 M_1}}}{\propto \frac{1}{\sqrt{m_2 M_2}}}
 {\propto \frac{\varrho}{\sqrt{m_2 M_3}}}{\propto \frac{1}{\sqrt{m_3 M_1}}}
 {\propto \frac{1}{\sqrt{m_3 M_2}}}{\propto \frac{\varrho}{\sqrt{m_3 M_3}}},
\end{equation}
where we have kept all $m_j$'s and $M_j$'s as variables while assuming $U_{e3}=0$ in deriving this. In the limit $\varrho\rightarrow 0$, one can immediately conclude that $\Omega_{23}, \Omega_{33} \rightarrow 0$ because $m_{2,3}$ and $M_3$ are finite and non-zero. On the other hand, this model predicts $m_1=0$ and therefore, at first glance, $\Omega_{13}$ is indeterminate while $\Omega_{11}$ and $\Omega_{12}$ are infinite as $m_1, \varrho\rightarrow 0$. But by the orthogonality condition, both issues can be resolved and $\Omega$ simplifies to
\begin{equation}\label{equ_omega3}
 \Omega=\3by3Mat{0}{0}{1}
                {\sqrt{1-\Omega_{31}^2}}{-\Omega_{31}}{0}
                {\Omega_{31}}{\sqrt{1-\Omega_{31}^2}}{0}.
\end{equation}
It is important to realise that this orthogonal see-saw matrix is identical to the special form derived from models with $M_3\rightarrow \infty$ (or effectively models with only two RH neutrinos) \cite{2RHnu_list, Ibarra:2003xp}.\footnote{This observation also justifies our claim that Case 1 and 3 are in fact equivalent.} The only distinction between these generic models and ours is that (\ref{equ_omega3}) is originated from the Yukawa couplings and family symmetry (which gives $m_1=0$ as a by-product). Therefore, the constraints from our Case 1 on the $CP$ asymmetry and their subsequent implications in thermal leptogenesis would be very similar to those models with only two RH neutrinos \cite{Ibarra:2003up}.\\

The first point we notice is that, in Case 1, $\m1eff$ is bounded from below. This can be illustrated by using (\ref{equ_omega3}) in the definition of $\m1eff$ and one obtains
\begin{equation}\label{equ_m1eff_case1}
 \m1eff = m_2 |1-\Omega_{31}^2|+m_3|\Omega_{31}^2|.
\end{equation}
When $|\Omega_{31}^2|\ll 1$, then $\m1eff\simeq m_2$ whereas when $|\Omega_{31}^2|\gg 1$, then $\m1eff\gg m_2$. Thus, we have $\m1eff \gtrsim m_2$. Notice that in the previous case, we can safely set $X_2=Y_2=X_3=0$ in order to maximise $\beta_2$ which leads to a relation $\m1eff = m_3 Y_{3\textrm{m}}$, and hence, no bounds on $\m1eff$. But in the current case, $X_2$ and $Y_2$ are no longer arbitrary as the freedom to change them while keeping $\m1eff$ constant has been lost due to the fact that $\Omega_{11}=0$. Because of this reduction of free variables, it is more straight forward to consider the function $\beta$ directly which is now modified to
\begin{equation}
 \beta =\frac{(m_3^2-m_2^2)\,Y_3}{\matm\,\m1eff}\,,\label{equ_beta_case1}
\end{equation}
with
\begin{equation}
 \m1eff =m_2\sqrt{(1-X_3)^2+Y_3^2}+m_3\sqrt{X_3^2+Y_3^2}\;,
\end{equation}
where we have imposed $X_2=1-X_3$ and $Y_2=-Y_3$. As a result of the bound: $\m1eff\gtrsim m_2$, the size of $\beta$ (and hence the $CP$ asymmetry) is very sensitive to the size of $Y_3$. This is because, unlike before where $Y_3\gg 1$  and $Y_3\ll 1$ lead to $\m1eff\gg1$ and $\m1eff\ll 1$ respectively, which then means that $\beta\; (\propto Y_3/\m1eff)$ approaches the same value ($\simeq 1$) for the two limiting cases, in (\ref{equ_beta_case1}) we have instead
\begin{equation}\label{equ_beta_limit}
 \beta = \frac{(m_3^2-m_2^2)}{\matm}\times
 \begin{cases}
  \displaystyle{\quad\frac{Y_3}{m_2}}\\
  \displaystyle{\frac{Y_3}{(m_3+m_2)\,Y_3}}
 \end{cases}
 \longrightarrow 
 \begin{cases}
  \quad 0 & \textrm{if}\; Y_3\ll X_3\ll 1,\\
  \displaystyle{\frac{m_3-m_2}{\matm}}& \textrm{if}\; Y_3\gg X_3\gg 1.
 \end{cases}
\end{equation}
Therefore in order to obtain maximum $CP$ asymmetry, $\m1eff$ must be very large. However, this is potentially detrimental to the success of thermal leptogenesis because $\m1eff$ controls the potency of the washout rates (hence the size of the final efficiency factor, see Sec.~\ref{subsec_eff}) and is therefore upper bounded if the correct $L$ asymmetry is to be generated. Fortunately, $M_1$ also dictates the final efficiency factor and the net result is that in order to circumvent the problem, one requires a larger lower bound for $M_1$ at the same time \cite{Buchmuller:2002rq}. Within the non-supersymmetric context, the mass of $M_1$, which is related to $T_\textrm{reh}$, is not tightly constrained and thus Case 1 of our model will be workable in leptogenesis.\\

Another observation from (\ref{equ_beta_limit}) is that the maximum attainable $\beta$ value in the limit of $Y_3\gg X_3\gg 1$ is drastically different depending on the hierarchy scheme assumed for the light neutrinos. For normal hierarchy, $\beta_\textrm{max}\approx 0.82$ whereas in the inverted case $\beta_\textrm{max}\approx 0.18$. Although this result alone is not sufficient to rule out the inverted case, it is clear that, for Case 1 at least, the inverted hierarchy is strongly disfavored.\\ 

Overall, we have demonstrated in this subsection that our specific neutrino models with family symmetry can naturally generate the required raw $CP$ asymmetry necessary for successful leptogenesis. The parameter space in these models is not overly restrictive, and the predictions are identical to a couple of special cases in the default see-saw, namely, the fully hierarchical light neutrinos limit (Case 2) and the two RH neutrinos scenario (Case 1 or 3).

\subsection{Efficiency factor}\label{subsec_eff}

The dynamical generation of a $B-L$ asymmetry in the leptogenesis era depends on the out-of-equilibrium decays of the heavy $N_j$'s, as well as other interactions in the thermal plasma. These non-equilibrium processes which control the evolution of $\curlN_{N_j}$ and $\curlN_{B-L}$ are quantified by a system of (at least) two Boltzmann kinetic equations. An important issue in setting up these is the identification of all relevant interactions which can modify $\curlN_{N_j}$ and $\curlN_{B-L}$ (see Table~\ref{table_4}). In the minimal setup where decays of the lightest RH neutrino $N_1$ dominate the final $B-L$ asymmetry and assuming hierarchical RH neutrino masses, only the evolution of $\curlN_{N_1}$ and $\curlN_{B-L}$ are relevant and the kinetic equations (in the default see-saw case) have the follow form \cite{Luty:1992un, Plumacher:1996kc, Buchmuller:2002rq, Barbieri:1999ma, Plumacher:1997ru}:
\begin{align}
 \odifone{\curlN_{N_1}}{z} &= -(D+S)(\curlN_{N_1}-\curlN_{N_1}^\textrm{eq}),\label{equ_BE1}\\
 \odifone{\curlN_{B-L}}{z} &= -\varepsilon_1\,D\,(\curlN_{N_1}-\curlN_{N_1}^\textrm{eq})
    -W\,\curlN_{B-L},\label{equ_BE2}
\end{align}
where the dimensionless variable $z=M_1/T$ is defined for convenience. $\curlN_{N_1}^\textrm{eq}$ denotes the equilibrium value for $\curlN_{N_1}$ which is now a function of $z$. The terms $D=\Gamma_D/(z H), S=\Gamma_S/(z H)$ and $W=\Gamma_W/(z H)$ encapsulate the reaction rate of the various processes with $H$ being the Hubble expansion rate which is given by
\begin{equation}
 H \simeq 1.66 \sqrt{g_s^*}\, \frac{M_1^2}{M_\textrm{Pl}}\frac{1}{z^2}\,,
\end{equation}
where $g_s^*(z\simeq 1) =106.75$ is the number of relativistic degrees of freedom,\footnote{This includes all SM particles only. The $N_1$ degrees of freedom is not included because in the preferred \textit{strong} washout regime, they are non-relativistic at $t_\textrm{lepto}$ (corresponding to $z_\textrm{lepto}=M_1/T_\textrm{lepto}\simeq 1$).} $M_\textrm{Pl}\approx 1.22\times 10^{19}$~GeV is the Planck mass. The thermally averaged (total) decay rate $\Gamma_D$ which accounts for decays and inverse decays ($N_1\leftrightarrow l\Phi$), is related to the zero-temperature decay rate $\Gamma_D^{(T=0)}\equiv\Gamma_1+\overline{\Gamma}_1$ via
\begin{equation}
 \Gamma_D = \Gamma_D^{(T=0)} \frac{\mathcal{K}_1(z)}{\mathcal{K}_2(z)}\,,
\end{equation}
where $\mathcal{K}_n(z)$ is the $n$th order modified Bessel function of the second kind. $\Gamma_S$ represents the $\Delta L =\pm 1$ scatterings (eg. $N_1 l\leftrightarrow \overline{t} q\,$). Typically, scatterings which involve gauge bosons $V_\mu$ are ignored to first approximation.\footnote{It turns out that they are only critical if one considers the \textit{weak} washout regime. See for example \cite{Giudice:2003jh, PilafUnderW, Covi:1997dr}} The washout rate $\Gamma_W$, which incorporates everything that tends to erase the $B-L$ asymmetry, is depended on the rates for inverse decay ($l\Phi\rightarrow N_1$), all $\Delta L=\pm 1$ scatterings (except those with $V_\mu$), as well as, the $\Delta L=\pm 2$ processes mediated by $N_1$ (eg. $l \Phi\leftrightarrow \overline{l} \Phi^\dagger$).\footnote{$\Delta L=\pm 2$ processes mediated by $N_{2,3}$ are suppressed at $z\simeq 1$ because $M_{2,3}\gg M_1$.}\\

\begin{table}[t]
\begin{center}
\begin{tabular}{|l||l|}
\hline
$\underline{\Delta L=0}$ & $\underline{\Delta L=\pm 1}$\\ 
$\quad H_\chi\leftrightarrow N_j N_j$ (tree + loop)$\star$
      & $\quad N_j\leftrightarrow l \Phi$ (tree + loop)\\
$\quad N_j N_j\leftrightarrow N_k N_k$ ($H_\chi, s$)$\star$
      & $\quad N_j\;l\leftrightarrow \overline{t}\; q$ ($\Phi, s$)\\
      & $\quad N_j\;t\leftrightarrow \overline{l}\; q$ ($\Phi, t$)\\
$\quad N_j N_j\leftrightarrow N_j N_j$ ($H_\chi, s$)$\star$
      & $\quad N_j\;q\leftrightarrow l\; t$ ($\Phi, t$)\\
$\quad N_j N_j\leftrightarrow N_j N_j$ ($H_\chi, t$)$\star$
      &\\
$\quad N_j N_k\leftrightarrow N_j N_k$ ($H_\chi, t$)$\star$
      & $\quad N_j\;\overline{l}\leftrightarrow \Phi\; V_\mu$ ($\Phi, s$)\\
      & $\quad N_j\;\overline{l}\leftrightarrow \Phi\; V_\mu$ ($l, t$)\\
$\quad N_j N_j\leftrightarrow H_\chi\Phi^\dagger\Phi$ ($H_\chi, s$)$\star$
      & $\quad N_j\, V_\mu\leftrightarrow \Phi\; l$ ($\Phi, t$)\\
$\quad N_j H_\chi\leftrightarrow N_j\Phi^\dagger\Phi$ ($H_\chi, t$)$\star$
      & $\quad N_j\, V_\mu\leftrightarrow \Phi\; l$ ($l, t$)\\
      & $\quad N_j\Phi^\dagger\leftrightarrow V_\mu\, l$ ($\Phi, t$)\\      
$\underline{\Delta L=\pm 2}$
      & $\quad N_j\Phi^\dagger\leftrightarrow V_\mu\, l$ ($l, s$)\\
$\qquad l\Phi\leftrightarrow \overline{l}\Phi^\dagger$ ($N_j, s$)
      &\\
$\qquad l\Phi\leftrightarrow \overline{l}\Phi^\dagger$ ($N_j, t$)
      & $\quad l\,\Phi\leftrightarrow N_j H_\chi$ ($N_j, s$)$\star$\\
$\qquad l\,l\leftrightarrow \Phi^\dagger\Phi^\dagger$ ($N_j, t$)
      & $\quad N_j\,l\leftrightarrow H_\chi \Phi^\dagger$ ($N_j, t$)$\star$\\
      & $\quad H_\chi\, l\leftrightarrow N_j \Phi^\dagger$ ($N_j, t$)$\star$\\
\hline
\end{tabular}\caption{A collection of potentially important processes in leptogenesis. The type of process (eg. tree-level, vertex/self-energy loop, $s$- or $t$-channel) and, where applicable, the mediating particle are in brackets. Interactions that are not in the default see-saw are marked by a ``$\star$''. $q$, $t$ and $V_\mu$ denote the (3rd generation) quark doublet, top quark singlet and gauge boson respectively.}
\label{table_4}
\end{center}
\end{table}

To describe the behavior of the solutions to (\ref{equ_BE1}) and (\ref{equ_BE2}), it is customary to introduce the decay parameter \cite{early_universe}
\begin{equation}\label{equ_decayK}
 K_1 = \frac{\Gamma_D^{(T=0)}}{H(z=1)}=\frac{\m1eff}{m_*}\,,
\end{equation}
where
\begin{equation}\label{equ_mstar}
 m_* =\frac{16}{3}\sqrt{\frac{\pi^5 g^*_s}{5}}\,\frac{\Dvev^2}{M_\textrm{Pl}} = \order{10^{-3}}\;
      \textrm{eV},
\end{equation}
is the equilibrium neutrino mass. $\m1eff$ is the effective neutrino mass as defined in (\ref{equ_m1eff}) which measures how strongly coupled $N_1$ is to the thermal plasma. It should be noted that the size of $\m1eff$ relative to $m_*$ marks the boundary between the so-called weak ($\m1eff < m_*$) and strong ($\m1eff > m_*$) washout regimes where the corresponding analyses are qualitatively different \cite{Buchmuller:2004nz}. This dependence of the solutions comes about because the interaction terms $D, S$ and $W_1$ (defined as $W_1 = W - \delta W$, where $\delta W$ represents the contribution from non-resonant $\Delta L=\pm 2$ processes) are proportional to $\m1eff$ \cite{Buchmuller:2002rq}:
\begin{equation}\label{equ_DSW_m1eff}
 D, S, W_1 \propto \frac{\Mpl\,\m1eff}{\Dvev^2}\,,\quad\textrm{while}\quad
 \delta W\propto \frac{\Mpl\, M_1\, \overline{m}^2}{\Dvev^4}\,,
\end{equation}
where $\overline{m}^2 = m_1^2+m_2^2+m_3^2$. For hierarchical light neutrinos and $M_1\ll 10^{14}$ GeV (note that both of these are satisfied in our models\footnote{The only restriction on $M_1$ is coming from (\ref{equ_T_rel}) which does not limit the possibility of $M_1\ll 10^{14}$ GeV.}), it turns out that contribution from $\delta W$ can be safely neglected \cite{Buchmuller:2002rq, Buchmuller:2004nz}, and therefore the generated $B-L$ asymmetry is to a good approximation independent of $M_1$. In the interesting case of \textit{strong} washout, the $W_1$ term is dominated by inverse decays \cite{Giudice:2003jh, Buchmuller:2004nz}. Moreover, in this regime, the heavy $N_1$'s in the plasma can reach thermal abundance before $t_\textrm{lepto}$ even if the scattering term $S$ is turned off. As a result, the details of $N_1$'s production prior to their decays become irrelevant and all subsequent analyses are greatly simplified. In the light of this, we shall concentrate on the strong washout regime in much of our discussions here.\\

With the above simplifications, (\ref{equ_BE1}) and (\ref{equ_BE2}) reduce to
\begin{align}
 \odifone{\curlN_{N_1}}{z} &= -D\,(\curlN_{N_1}-\curlN_{N_1}^\textrm{eq}),\label{equ_BE3}\\
 \odifone{\curlN_{B-L}}{z} &= -\varepsilon_1\,D\,(\curlN_{N_1}-\curlN_{N_1}^\textrm{eq})
    -W_1^\textrm{ID}\curlN_{B-L},\label{equ_BE4}
\end{align}
where $W_1^\textrm{ID}(	\simeq W_1)$ is the dominant piece in the washout term which originates from inverse decays.\footnote{Note that in (\ref{equ_BE3}) and (\ref{equ_BE4}), the resonant $\Delta L=\pm 2$ contribution must be properly accounted for to avoid the un-physical asymmetry generation in thermal equilibrium \cite{Giudice:2003jh, KolbWolf_Dolgov}.} The solution for $\curlN_{B-L}$ can be expressed in an integral form \cite{Buchmuller:2004nz, Kolb:1983ni}
\begin{equation}\label{equ_soln1}
 \curlN_{B-L}(z) = \curlN_{B-L}^{\textrm{i}}\,e^{-\int_{z_\textrm{i}}^z dz'\, W_1^\textrm{ID}(z')} 
  -\varepsilon_1\,\kappa_1(z),
\end{equation}
where $\curlN_{B-L}^{\textrm{i}}$ and $z_\textrm{i}$ denote the initial values for $\curlN_{B-L}$ and $z$ respectively. The corresponding efficiency factor is given by
\begin{equation}\label{equ_kappa1}
 \kappa_1(z)=-\int_{z_\textrm{i}}^{z}\,dz'\,\odifone{\curlN_{N_1}}{z'}\, 
   e^{-\int_{z'}^{z}\,dz''\,W_1^\textrm{ID}(z'')}.
\end{equation}
Since in the strong washout regime one can invoke the approximation $d\curlN_{N_1}/dz'\simeq d\curlN_{N_1}^\textrm{eq}/dz'$, the final value as $z\rightarrow\infty$ may be readily worked out and one obtains \cite{Buchmuller:2004nz}
\begin{equation}\label{equ_kappa2}
 \kappa^\textrm{f}_1(K_1)\simeq 
   \frac{2}{K_1\,z_B(K_1)}\left(1-e^{-\frac{K_1\,z_B(K_1)}{2}}\right),\quad\textrm{for } K_1\gtrsim 1,
\end{equation}
where we have expressed $\kappa^\textrm{f}_1$ as a function of the decay parameter $K_1$; and $z_B$ ($\gg z_\textrm{lepto}\simeq 1$) is the value around which the asymptotic expansion of the $z'$ integral in (\ref{equ_kappa1}) receives a dominant contribution. The temperature $T_B$ that corresponds to $z_B$ ($\equiv M_1/T_B$) is referred to as the \textit{baryogenesis temperature}. For practical purposes, (\ref{equ_kappa2}) is well approximated by the simple power law \cite{Buchmuller:2004nz}:
\begin{equation}\label{equ_kappa3}
 \kappa^\textrm{f}_1 \simeq (2\pm 1)\times 10^{-2}\left[
    \frac{0.01\,\textrm{eV}}{\m1eff}\right]^{1.1\pm 0.1},
   \quad\textrm{for}\; \m1eff > m_*.
\end{equation}

Assuming we are working in the strong washout region, we will now argue that the standard results described above are directly applicable to our models with family symmetry. Of the $\Delta L =0$ or $\Delta L =\pm 1$ new interactions originating from coupling to $H_\chi$, all involving an external $H_\chi$ in the initial state (eg. $H_\chi\rightarrow N_j N_j$) can be disregarded because the reaction density for $H_\chi$ is almost zero at $z\simeq 1$ while the reverse processes are kinematical forbidden due to the heaviness of $H_\chi$.\footnote{It should be noted that the $2\leftrightarrow 3$ reactions such as $N_j N_j\leftrightarrow H_\chi\Phi^\dagger\Phi$ are in any case impotent because of $\mu_{\phi\chi}\rightarrow 0$.} Furthermore, the $N_j$-$N_k$ scatterings mediated by a virtual $H_\chi$ do not play a role because in the \textit{strong} washout regime, the analysis is insensitive to the initial $N_j$ abundance in the thermal plasma. Hence, there is no new important contribution from the list of starred ($\star$) reactions in Table~\ref{table_4} to the standard scattering and washout terms, and we can conclude that our models do not predict a modification to the efficiency factor $\kappa_1^\textrm{f}$ given by the default see-saw case.

\subsubsection*{Weak washout regime}

We end this section by including a short discussion of the phenomenologies that might be essential to the analysis of our extended see-saw models in the context of weak washout.\footnote{For a complete review of the analysis in the weak washout regime, see for example \cite{Giudice:2003jh, Buchmuller:2004nz}.} While the efficiency factor is typically enhanced in this regime, one obvious drawback is that the prediction is no longer independent of initial conditions as in the scenario for strong washout. As a result, the $N_j$-$N_k$ scatterings via a $H_\chi$ may be significant since these can actively modify $\curlN_{N_1}^{\textrm{i}}$ in the plasma while inverse decays are no longer strong enough to ensure thermal abundance before the onset of leptogenesis.\\

However, upon closer inspection and assuming a hierarchial RH neutrino mass spectrum, one would not expect a sizable change to $\curlN_{N_1}^{\textrm{i}}$. This is because in Table~\ref{table_4} there is no ($\Delta L =0$) creation or annihilation process which involves $N_1$'s and some \textit{lighter} particles. Note that one does not need to be concerned with $\Delta L \neq 0$ processess in this analysis because they are in general too weak to bring $N_j$'s into equilibrium at high temperature ($T_\textrm{lepto} \ll T < T_\textrm{reh}$). Thus only processes such as $N_1 N_1 \leftrightarrow N_k N_k\; (k=2,3)$ can bring $N_1$'s into equilibrium. But as has been shown in a similar model in \cite{Plumacher:1996kc} where the additional $N_j$-$N_k$ interactions come from couplings to massive neutral gauge bosons (related to GUT breaking), these scatterings have very little effect on the final asymmetry prediction if one has a pronounced RH neutrino mass hierarchy.\footnote{The model presented in \cite{Plumacher:1996kc} provides a good guide to the phenomenologies expected in ours with the exception that our models do not possess interactions that link $N_j$ to SM particles. Consequently, our models would predict even less modifications than in the case of \cite{Plumacher:1996kc}.} In addition, the $N_1$-dominated approximation remains valid.\\

It should be pointed out that all interactions involving an external $H_\chi$ remain unimportant. When $z\ll 1$, processes that can produce $N_j$ (eg. $H_\chi\rightarrow N_j N_j$) are irrelevant because they are only potent before the inflationary stage (cf. (\ref{equ_T_rel})) and any excess $N_j$ created will be diluted away. On the other hand, for $z\simeq 1$, reaction density for $H_\chi$ is almost zero as in the strong washout regime.\\

Other issues that are potentially important in the weak washout regime include the effects of scattering processes that couple to gauge bosons $V_\mu$  \cite{Giudice:2003jh, PilafUnderW} (see Table~\ref{table_4}), as well as thermal corrections to $\Gamma_D$ and $\Gamma_S$ \cite{Giudice:2003jh, Covi:1997dr}. These elements provides an additional source of theoretical uncertainty to the overall analysis and hence making the weak washout scenario less attractive than the strong washout case. For the purpose of this paper though, these effects unnecessarily complicate the analysis and therefore will be ignored.\\

Although we have not shown by explicit calculations that predictions of our models will not significantly deviate from (or bias towards certain parameter space within) the standard case with weak washout, it is apparent from the above discussion that a lot of it will be highly dependent on the assumptions made on the mass of $N_j$, $H_\chi$ and their Yukawa couplings. Given that all of these are free parameters in our models, any deviations with respect to the standard see-saw case will therefore stay within the amount of these uncertainties. Hence, we can conclude that their predictions are effectively the same as the default case.


\section{Implications in leptogenesis with flavor effects}\label{sec_lep2}

In analogy to the discussion in Sec.~\ref{subsec_dilution} on spectator processes, flavor effects become important when the assumption on the temperature range in which leptogenesis happens\footnote{In the $N_1$-dominated scenario, this corresponds to $T\simeq M_1$. The moral is that if $M_1$ is light enough then flavor effects become important.} is altered \cite{Barbieri:1999ma, Endoh:2003mz, Fujihara:2005pv,Vives:2005ra, Nardi:2006fx, Abada:2006fw, Abada:2006ea, Blanchet:2006be}. This arises because Yukawa interactions become more prominent at temperatures $T\lesssim 10^{12}$ GeV. Whereas in the previous discussion the role of these interactions were merely to change the reaction densities in the Boltzmann equations which led to what we referred to as a new effective sphaleron conversion factor,\footnote{One may also simply interpret it as another form of modification to the washout terms.} in the context of flavor issues in leptogenesis, their effects on the $CP$ asymmetry and (flavor-dependent) washout are also considered.\\

Whilst in the one-flavor approximation, the lepton states, $|\,l_{(j)}\rangle$'s,\footnote{The subscript $(j)$ highlights the fact that the flavor decomposition of $l_{(j)}$ can be different for each $j$.} generated by the decays of $N_j$'s are assumed to have evolved \textit{coherently} during the leptogenesis era, for the flavor-aware situation, the presence of charged lepton Yukawa interactions in equilibrium essentially introduces a source of decoherence whereby the $|\,l_{(j)}\rangle$'s are projected onto one of the three flavor eigenstates, $|\,l_{\alpha}\rangle$'s ($\alpha = e, \mu, \tau$) with probability $|\langle l_{(j)}|\,l_{\alpha}\rangle|^2$. Because of the difference in size between the tauon and muon Yukawa couplings, there exists two temperature thresholds: $T^\textrm{eq}_\tau$ and $T^\textrm{eq}_\mu$, which govern the range where tauon and muon interactions come into equilibrium respectively. When $T > T^\textrm{eq}_\tau\simeq 10^{12}$~GeV, both the $\tau$- and $\mu$-Yukawa interactions are out-of-equilibrium and flavor effects can be ignored. For the temperature range $T^\textrm{eq}_\tau \gtrsim T\gtrsim T^\textrm{eq}_\mu\simeq 10^{9}$~GeV \cite{Barbieri:1999ma, Abada:2006fw, Abada:2006ea}, only $\tau$-Yukawas are in equilibrium and one effectively has a two-flavor ($\tau$ and a linear combination of $\mu$ and $e$) problem. Finally, in the case of $T^\textrm{eq}_\mu \gtrsim T$, both reactions are strong enough to instigate a full three-flavor system. In our discussion here, we do not explicitly distinguish between the two situations and simply assume the three-flavor regime. Furthermore, spectator effects in the form discussed in \cite{spectator1, spectator2} will be ignored for brevity.\\

To illustrate the origin of the flavor affected $CP$ asymmetry,\footnote{A proper discussion of this should be done within the density matrix framework \cite{Barbieri:1999ma, Abada:2006fw}. But for our purpose, it is enough to follow the more intuitive approach as in \cite{Nardi:2006fx}.} we begin by rewriting  Lagrangian (\ref{equ_lag2}) with sub-indices
\begin{equation}\label{equ_lag3}
 \mathcal{L}_{\textrm{mass}} = -\frac{1}{2}\overline{N}_j\,D_{M_j}\,N_j
                               -\overline{l}_\alpha\,(h_l)_\alpha\,\Phi\, e_\alpha
                               -\overline{l}_\alpha\,(h_\nu)_{\alpha j}\,\widetilde{\Phi}\,N_j 
                               +\textrm{h.c.},
\end{equation}
where $j=1,2,3$ and $\alpha=e,\mu,\tau$. It is then clear that the state $|\,l_{(j)}\rangle$ and $|\,\overline{l}_{(j)}\rangle$ in (\ref{equ_cp1}) are in general \textit{not} a $CP$ conjugate of each other owning to the complex entries in matrix $h_\nu$. If we are considering the temperature ranges where the relevant lepton Yukawa interactions are either fully in equilibrium or out-of-equilibrium but not in bewteen such that state $|\,l_{(j)}\rangle$ will  quickly decohere into one of the $|\,l_{\alpha}\rangle$ states avaliable, then we can effectively think of $\Gamma_j$ as an incoherent sum over $\alpha$ of the partial decay rates
$\Gamma_{j\alpha}\equiv \Gamma(N_j\rightarrow l_\alpha \Phi)$. Likewise for $|\,\overline{l}_{(j)}\rangle, |\,\overline{l}_\alpha\rangle, \overline{\Gamma}_j$ and $\overline{\Gamma}_{j\alpha}\equiv\Gamma(N_j\rightarrow \overline{l}_\alpha\Phi^\dagger)$. Note here that $|\,l_{\alpha}\rangle$ and $|\,\overline{l}_\alpha\rangle$ are $CP$ conjugate states of each other. It is then straightforward to see that $CP$ violation in $N_j$ decays can manifest itself in two places:
\begin{enumerate}
 \item[A.] The amount of $|\,l_\alpha\rangle$ and $|\,\overline{l}_\alpha\rangle$ produced are not the same because $|\,l_{(j)}\rangle$ and $|\,\overline{l}_{(j)}\rangle$ are produced at different rates which corresponds to $\Gamma_j\neq \overline{\Gamma}_j$. 
 \item[B.] The amount of $|\,l_\alpha\rangle$ and $|\,\overline{l}_\alpha\rangle$ produced are not the same because $\Gamma_{j\alpha}\neq \overline{\Gamma}_{j\alpha}$ (regardless of the relation between $\Gamma_j,  \overline{\Gamma}_j$).
\end{enumerate}
%


%
%

Obviously, the second effect is only relevant when one is considering the evolution of the individual lepton flavor asymmetry $L_\alpha$. For the one-flavor approximation, one tracks the evolution of $L=\sum_\alpha L_\alpha$ instead and hence only the first effect comes into play. However, an important corollary is that when flavor effects are included, the associated \textit{flavored} $CP$ asymmetry, $\varepsilon_{j\alpha}$, can be non-zero even if $\Gamma_j =\overline{\Gamma}_j$ (ie. $\varepsilon_j =0$).\footnote{This situation corresponds to having a real orthogonal matrix $\Omega$ (cf. (\ref{equ_omega_1})) which in turn implies that $CP$ is an exact symmetry in the RH neutrino sector \cite{Abada:2006ea}. An intriguing consequence of this situation is that leptogenesis is directly connected to the low energy $CP$ violating phases in mixing matrix $U$ \cite{lowCP_lepto_recent}.} To properly quantify all these, it is convenient to introduce the flavor projectors \cite{Nardi:2006fx}
\begin{align}
 P_{j\alpha} &\equiv \frac{\Gamma_{j\alpha}}{\Gamma_{j}} = |\langle l_{(j)}|\,l_{\alpha}\rangle|^2\,,
 \label{equ_P1}\\
 \qquad\qquad\qquad\qquad\qquad\qquad
 \overline{P}_{j\alpha} &\equiv \frac{\overline{\Gamma}_{j\alpha}}{\overline{\Gamma}_{j}} 
 = |\langle \overline{l}_{(j)}|\,\overline{l}_{\alpha}\rangle|^2\,,\qquad
  (j=1,2,3;\, \alpha=e,\mu,\tau)
 \label{equ_P2}
\end{align}
where $\Gamma_j =\sum_\alpha\, \Gamma_{j\alpha}$ and $\overline{\Gamma}_j =\sum_\alpha\, \overline{\Gamma}_{j\alpha}$. Therefore, by definition, $\sum_\alpha\, P_{j\alpha} =\sum_\alpha\, \overline{P}_{j\alpha}=1$. The associated $\alpha$ flavor $CP$ asymmetry is given by
\cite{Barbieri:1999ma, Nardi:2006fx}
\begin{align}
 \varepsilon_{j\alpha} \equiv \frac{\Gamma_{j\alpha}-\overline{\Gamma}_{j\alpha}}
                       {\Gamma_{j}+\overline{\Gamma}_j}
                       &=\frac{\Gamma_{j}P_{j\alpha}-\overline{\Gamma}_j\overline{P}_{j\alpha}}
                       {\Gamma_{j}+\overline{\Gamma}_j}\,,\label{equ_fcp1}\\
                       &=\frac{P_{j\alpha}+\overline{P}_{j\alpha}}{2}\,\varepsilon_j
                        +\frac{P_{j\alpha}-\overline{P}_{j\alpha}}{2}\,,
                       \label{equ_fcp2}\\
                       &\simeq P^0_{j\alpha}\,\varepsilon_j+\frac{\delta P_{j\alpha}}{2}\,,
                       \label{equ_fcp3}
\end{align}
where $P^0_{j\alpha}$ is the tree level contribution to the projector $P_{j\alpha}$ with $P^0_{j\alpha}=\overline{P}^0_{j\alpha}$, while $\delta P_{j\alpha}\equiv P_{j\alpha} -\overline{P}_{j\alpha}$ is the quantity that characterises the $CP$ violating effect of type B. From (\ref{equ_fcp3}), it is clear that even if $\varepsilon_j=0$ and hence $P^0_{j\alpha}\,\varepsilon_j$ (representing type A effect) vanishes, $\varepsilon_{j\alpha}$ does not necessarily go to zero. In addition, 
\begin{equation}
 \sum_\alpha\,\delta P_{j\alpha} = \sum_\alpha\,(P_{j\alpha} -\overline{P}_{j\alpha})
  = 1- 1=0,
\end{equation}
demonstrating that $\varepsilon_j\equiv \sum_\alpha\,\varepsilon_{j\alpha}$ will not depend on $\delta P_{j\alpha}$, and hence it is consistent with our claim that only type A effect is important in the one-flavor approximation. Putting these quantities in terms of parameters in (\ref{equ_lag3}) and setting $j=1$ for the $N_1$-dominated scenario, we get \cite{Nardi:2006fx}
\begin{align}
  P_{1\alpha}^0 =\overline{P}^0_{1\alpha} &= \frac{(h_\nu^*)_{\alpha 1}(h_\nu)_{\alpha 1}}
  {(h_\nu^\dagger h_\nu)_{11}}\,,\label{equ_flavourP0a}\\
  &=\frac{|\sum_k\,\sqrt{m_k}\,U_{\alpha 1} \Omega_{k1}|^2}{\sum_k\,m_k |\Omega^2_{k1}|}\,,
  \label{equ_flavorP0}
\end{align}
and \cite{Covi:1996wh}
\begin{equation}\label{equ_flavorCP}
 \varepsilon_{1\alpha}=-\frac{1}{8\pi(h_\nu^\dagger h_\nu)_{11}}
   \sum_{k\neq 1}\textrm{Im}\left\{ (h_\nu^*)_{\alpha 1}(h_\nu)_{\alpha k}\frac{M_1}{M_k}
   \left[\,\frac{3}{2}(h_\nu^\dagger h_\nu)_{1k}+\frac{M_1}{M_k}(h_\nu^\dagger h_\nu)_{k1}\right]
   \right\}\,,
\end{equation}
with $\varepsilon_1$ given by (\ref{equ_cp3a}). Using these definitions, one can then calculate $\delta P_{1\alpha}$ via (\ref{equ_fcp3}).\footnote{Even without explicitly calculating $\delta P_{1\alpha}$, one can see from (\ref{equ_flavorCP}) that the term proportional to $(h_\nu^\dagger h_\nu)_{k1}$ drops out upon summing over $\alpha$, since when combined with the pre-factor outside the brackets, it becomes $(h_\nu^\dagger h_\nu)_{1k}(h_\nu^\dagger h_\nu)_{k1}$ which is real. This is an indication that this term actually corresponds to $\delta P_{1\alpha}$ type effect.}\\

To derive the network of Boltzmann equations, firstly we note that the final asymmetry $\curlN^\textrm{f}_{B-L}$ would now depend on the evolution of all individual flavor asymmetries $\curlN_{Q_\alpha}$ where $Q_\alpha \equiv B/3-L_\alpha$ and $\curlN^\textrm{f}_{B-L}=\sum_\alpha\,\curlN^\textrm{f}_{Q_\alpha}$. Secondly, the washout terms would be modified in the multi-flavor case because $\Delta L \neq 0$ interactions with $\Phi$ couple to state $|\,l_{(1)}\rangle$ and not $|\,l_\alpha\rangle$. This effect is accounted for by $P_{1\alpha}\simeq P_{1\alpha}^0$.  If we assume again that the Yukawa interactions are either strongly in equilibrium or out-of-equilibrium but not in the transition region, then the kinetic equations are greatly simplified as flavor dynamics due to the coherences may be neglected.\footnote{These coherences, in the density matrix formulism \cite{Barbieri:1999ma}, correspond to the off-diagonal terms, $\rho_{\alpha\beta}$ of a density operator, $\rho$, which has the properties: $\rho_{\alpha\alpha}\propto \curlN_{L_\alpha}$ and $\sum_\alpha\,\rho_{\alpha\alpha}=\curlN_{L}$. In cases where these effects cannot be ignored, one must analyse the system in full. See for example \cite{Abada:2006fw}.} Furthermore, in analogy to Sec.~\ref{subsec_eff}, a good approximation to the problem can be obtained by considering just the effect of decays and inverse decays\footnote{The real intermediate state contribution from $\Delta L=\pm 2$ scatterings has been properly subtracted.} \cite{Blanchet:2006be}. Overall, we get (cf. (\ref{equ_BE3}) and (\ref{equ_BE4})) \cite{Nardi:2006fx, Blanchet:2006be}
\begin{align}
 \odifone{\curlN_{N_1}}{z} &= -D\,(\curlN_{N_1}-\curlN_{N_1}^\textrm{eq}),\label{equ_fBE1}\\
 \odifone{\curlN_{Q_\alpha}}{z} &= -\varepsilon_{1\alpha}\,D\,(\curlN_{N_1}-\curlN_{N_1}^\textrm{eq})
    -P^0_{1\alpha}\,W_1^\textrm{ID}\,\curlN_{Q_\alpha}.\label{equ_fBE2}
\end{align}
Like (\ref{equ_soln1}), the solution to $\curlN_{Q_\alpha}$ can be expressed in an integral form:
\begin{equation}\label{equ_fsoln1}
 \curlN_{Q_\alpha}(z) = 
 \curlN_{Q_\alpha}^{\textrm{i}}\,e^{-P^0_{1\alpha}\int_{z_\textrm{i}}^z dz'\, W_1^\textrm{ID}(z')} 
  -\varepsilon_{1\alpha}\,\kappa_{1\alpha}(z),
\end{equation}
where the flavor dependent efficiency factor is given by
\begin{equation}\label{equ_fkappa1}
 \kappa_{1\alpha}(z)=-\int_{z_\textrm{i}}^{z}\,dz'\,\odifone{\curlN_{N_1}}{z'}\, 
   e^{-P^0_{1\alpha}\int_{z'}^{z}\,dz''\,W_1^\textrm{ID}(z'')}.
\end{equation}
In the strong washout regime, one then gets (cf. {\ref{equ_kappa2}) \cite{Blanchet:2006be}
\begin{equation}\label{equ_fkappa2}
 \kappa^\textrm{f}_{1\alpha}(K_{1\alpha})\simeq 
   \frac{2}{K_{1\alpha}\,z_B(K_{1\alpha})}
   \left(1-e^{-\frac{K_{1\alpha}\,z_B(K_{1\alpha})}{2}}\right),
\end{equation}
where $K_{1\alpha}\equiv P^0_{1\alpha} K_{1}$. Since $P^0_{1\alpha}\leq 1$, we have $K_{1\alpha} \leq K_{1}$ and so (\ref{equ_fkappa2}) implies that washout is in general reduced because of flavor effects. As a consequence of this, one finds that the region of parameter space corresponding to the strong washout regime is also reduced \cite{Blanchet:2006be}.\\

From the above discussion, it is clear that much of the analysis done for our models in Sec.~\ref{sec_lep1} would be applicable to the case when flavor effects are included. Comparing (\ref{equ_fBE1}) and (\ref{equ_fBE2}) with (\ref{equ_BE3}) and (\ref{equ_BE4}), we observe that $P^0_{1\alpha}$ and $\varepsilon_{1\alpha}$ are the only two new ingredients which contain all the additional model dependent information originating from flavor effects. Hence, it suffices to investigate the implications of our specific models on these two quantities.\\

To begin with, we note that both $P^0_{1\alpha}$ and $\varepsilon_{1\alpha}$ depends explicitly on the mixing matrix $U$, unlike in the one-flavor approximation where everything can be written in terms of the see-saw orthogonal matrix $\Omega$. This can be seen from (\ref{equ_flavorP0}) and by expanding the two terms containing $h_\nu$'s in (\ref{equ_flavorCP}) using (\ref{equ_omega1}):
\begin{equation}\label{equ_fCPterm1}
 (h_\nu^*)_{\alpha 1}(h_\nu)_{\alpha k}(h_\nu^\dagger h_\nu)_{1k}
 = \frac{M_1 M_k}{\Dvev^4}
    \sum_n \, m_n \Omega^*_{n1}\Omega_{nk}\,
    \sum_{\ell,\,m}\sqrt{m_\ell\,m_m}\: \Omega^*_{\ell 1}\Omega_{mk}U^*_{\alpha\ell}U_{\alpha m},
\end{equation}
and
\begin{equation}\label{equ_fCPterm2}
 (h_\nu^*)_{\alpha 1}(h_\nu)_{\alpha k}(h_\nu^\dagger h_\nu)_{k1}
 = \frac{M_1 M_k}{\Dvev^4}
    \sum_n \, m_n \Omega_{n1}\Omega^*_{nk}\,
    \sum_{\ell,\,m}\sqrt{m_\ell\,m_m}\: \Omega^*_{\ell 1}\Omega_{mk}U^*_{\alpha\ell}U_{\alpha m}.
\end{equation}
So, this illustrates that low energy $CP$ violating phases may, in prinicple, have direct connection to leptogenesis (with flavor effects). However, these effects are usually masked by the complex phases in $\Omega$ (ie. from $h_\nu$) and would enter directly only when $\Omega$ is real as stated before. In the context of our models with family symmetry, since the prediction is $U_{e3}=0$, the Dirac phase $\delta$ in (\ref{equ_PMNS}) does not enter into the theory at all, and hence $CP$ violating effects arising from low energy parameters would entirely come from the \textit{Majorana} phases $\varphi_{1,2}$.\\

In terms of whether our models would give rise to a significant departure in the predictions of leptogenesis (with flavors), there are some general observations we can make.\footnote{Note that there is nothing in all cases of our models which specifically indicates whether flavor effects should be included or not for $M_1$ is a free parameter in the theory. Thus, leptogenesis with flavor effects is phenomenologically not excluded by our models.} Firstly, by examining (\ref{equ_flavourP0a}) and (\ref{equ_flavorCP}), we can safely conclude that for Case 2 of our models (see Table~\ref{table_3}) there is essentially no modifications to the standard see-saw scenario. This is due to the fact that all entries in $h_\nu$ for this case are uncontrained by the symmetry and therefore it can, in principle, accommodate all specific scenarios the default see-saw allows (except $m_1=0$ must be obeyed). Secondly, using expression (\ref{equ_flavorP0}) for $P^0_{1\alpha}$, we see that both Case 1 and 2 do not provide any predictions or restrictions on the projectors $P^0_{1\alpha}$'s.\footnote{Again Case 3 is automatically accounted for when considering Case 1.} This is because $P^0_{1\alpha}$ depends only on the 21- and 31-entries\footnote{No dependence on the 11-entry because $m_1=0$ and contribution to the sum in (\ref{equ_flavorP0}) automatically disappears.} of both $U$ and $\Omega$ and we know from before that these are unconstrained by the symmetry (for Case~1, see (\ref{equ_omega3})).\\

To study the implications of Case~1 on $\varepsilon_{1\alpha}$, we can apply $m_1=0, U_{e3}=0$ and $\Omega$ in (\ref{equ_omega3}) to expressions (\ref{equ_fCPterm1}) and (\ref{equ_fCPterm2}). A key feature which results is that both (\ref{equ_fCPterm1}) and (\ref{equ_fCPterm2}) vanishes when $k=3$, which is in accordance with the previous conclusion that Case~1 corresponds to the see-saw case with only two RH neutrinos. In fact, this is the only definite prediction from the model. Although one gets significant simplication to (\ref{equ_fCPterm1}) and (\ref{equ_fCPterm2}) for particular values of $k$ and $\alpha$, the resulting expressions are still a function of unconstrained parameters, hence one has to impose additional conditions in order to restrict the result. For instance, when $k=2$ and $\alpha=e$, (\ref{equ_fCPterm2}) simplifies to
\begin{equation}\label{equ_fterm1}
 (h_\nu^*)_{e 1}(h_\nu)_{e 2}(h_\nu^\dagger h_\nu)_{21}
 = \frac{m_2|U_{e2}|^2M_1 M_2}{\Dvev^4}\, (m_2 |\Omega_{31}^2| |1-\Omega_{31}^2|-m_3  \Omega_{31}^2(1-\Omega^2_{31})^*),
\end{equation}
where we have used $m_1=0, U_{e3}=0$ and (\ref{equ_omega3}). Plugging this into (\ref{equ_flavorCP}) and defining $\Omega_{31}^2=X_3+iY_3$, we simply get (ignoring $\order{(M_1/M_2)^2}$ terms):
\begin{equation}
 \varepsilon_{1e} \simeq -\frac{3 M_1}{16\pi \Dvev^2\, \m1eff}\, m_2^2\,|U_{e2}|^2\, Y_3. 
\end{equation}
This example illustrates that although the texture zeros can lead to partial simplification, in the end, one has to specify $\Omega_{31}^2$ and $U_{e2}$ (in this case) in order to make any predictions about $\varepsilon_{1e}$. Besides, the final asymmetry would in general depends on all of the $\varepsilon_{1\alpha}$'s and not just $\varepsilon_{1e}$. Therefore, all we may conclude is that Case~1 of these models will possess the phenomenologies given by the standard see-saw with two RH neutrinos. An extensive study on this particular situation in the context of leptogenesis with flavor effects has been performed recently in \cite{Abada:2006ea}.\\

So far, we have assumed the strong washout scenario exclusively. The reason for that is in the weak washout regime, the calculation of the final asymmetry relies on a good knowledge of the thermal history of all RH neutrinos in the plasma. Since our models do not provide specific information on that front and flavor effects cannot actually alter the phenomena we discussed in Sec.~\ref{subsec_eff}, therefore, our conclusion on this regard would be similar. In the light of all these, it is clear that implications of our models in leptogenesis with flavor effects are basically the same as those from the normal see-saw (with $m_1=0$).


\section{Summary of results}\label{sec_sum}

It is fascinating that two seemingly unrelated problems---the tiny masses of light neutrino and the cosmic baryon asymmetry---may be explained in a consistent manner by the mere introduction of heavy RH neutrinos to the SM. While the former may be explained by the see-saw mechanism, thermal leptogenesis provides an attractive solution to the later. Given this connection between the two, it is natural to ask whether a specific class of constrained see-saw models which has less free parameters than the default setup, can give rise to successful leptogenesis. In this paper, we have performed a thorough check in this context on such constrained models as those proposed by Low \cite{Low:2005yc}. These models obey certain Abelian family symmetry and contain one additional singlet in the Higgs sector such that the total number of arbitrary parameters in the see-saw theory is reduced. In particular, they predict that $\theta_{13}=0$ and a fully hierarchical light neutrino spectrum ($m_1=0$).\\

By dissecting the dependence of the final asymmetry on sphalerons, $CP$ violating decays and washouts, we have identified the key elements that can modify the leptogenesis predictions. Consequently, the implications of different see-saw neutrino models can be easily compared. It was found that in all cases of our models, leptogenesis predictions are almost identical to those allowed by the default see-saw model, with the exception that our model naturally selects the hierarchical light neutrino solution (for $m_1=0$ is one of its main features). This conclusion is true respectively for both the one-flavor approximation, as well as when flavor effects are included since all of the essential elements that can change the final asymmetry turn out to be unconstrained in our models. In one case, the phenomenologies are very much the same as the standard situation with three RH neutrinos whereas in the two other cases, they correspond to the scenario with only two RH neutrinos. Furthermore, for leptogenesis with flavor effects, we have found that Majorana phases in the light neutrino mixing matrix can play an important role since in our models the Dirac phase disappears due to the fact that $\theta_{13}=0$.\\

It should be noted that the entire investigation has been done assuming the limit of hierarchical RH neutrinos. Also, we have adopted the $N_1$-dominated scenario in most situations. Although in much of this paper the analyses are done in the strong washout regime, there are clear indications that our general conclusion can be extended to the weak washout case. A more precise calculation on this front, however, cannot be carried out since our models do not provide any specific predictions on the RH neutrino sector. On one hand, this means that nothing new is predicted by our models, but on the other, this non-existence of heavy constraints ensures that these models lead to successful leptogenesis in most scenarios.\\

\textbf{Note added:--}
While this work was in preparation, it was found in \cite{Blanchet:2006ch} that quantum Zeno effects can affect the conditions in which flavor effects in leptogenesis are applicable. Although this is an important observation, it will not alter equations (\ref{equ_fBE1}) and (\ref{equ_fBE2}). Hence, our analysis in Sec.~\ref{sec_lep2} and its conclusion remain the same.


\section*{Acknowledgments}

SSCL would like to sincerely thank S. Blanchet, P. Di Bari and C. Low for helpful hints and comments on the subject of this paper. This work was supported in part by the Australian Research Council and in part by the Commonwealth of Australia.

\end{document}